\documentclass[a4paper,11pt]{article}
\usepackage{jinstpub} 
\usepackage{booktabs}
\usepackage{lineno}
\usepackage{xspace}
\usepackage{ifthen}
\usepackage{upgreek}
\usepackage{subcaption}

\newcommand{\nospaceunit}[1]{\ensuremath{\text{#1}}}       
\newcommand{\aunit}[1]{\ensuremath{\text{\,#1}}}  
\def\lhcb   {\mbox{LHCb}\xspace}
\def\lhc    {\mbox{LHC}\xspace}
\def\cm   {\aunit{cm}\xspace}
\def\mm   {\aunit{mm}\xspace}
\def\sec  {\ensuremath{\aunit{s}}\xspace}
\def\pt         {\ensuremath{p_{\mathrm{T}}}\xspace}
\def\PK      {\ensuremath{\mathrm{K}}\xspace}   
\def\Kz      {{\ensuremath{\kaon^0}}\xspace}
\def\PLambda     {\ensuremath{\Lambda}\xspace}     
\def\kaon    {{\ensuremath{\PK}}\xspace}
\def\Lz          {{\ensuremath{\PLambda}}\xspace}
\def\mum  {\ensuremath{\,\upmu\nospaceunit{m}}\xspace}
\def\pt         {\ensuremath{p_{\mathrm{T}}}\xspace}
\def\p         {\ensuremath{p}\xspace}
\def\tbps        {\aunit{Tbit/s}\xspace}
\newcommand{\gevc}{\ensuremath{\aunit{Ge\kern -0.1em V\!/}c}\xspace}
\newcommand{\chisq}{\ensuremath{\chi^2}\xspace}

\title{\boldmath  Tracking performance study of the \lhcb UP Detector}

\author[a,b]{Yisheng Fu}
\author[a]{, Jianchun Wang}
\author[a]{, Xuhao Yuan}
\author[]{, on behalf of the LHCb UP team}
\affiliation[a]{Institute of High Energy Physics, Chinese Academy of Science, 19B Yuquan Road, Beijing, 100049, Beijing, China}
\affiliation[b]{University of Chinese Academy of Science, 19A Yuquan Road, Beijing, 100049, Beijing, China}

\emailAdd{xhyuan@ihep.ac.cn,jwang@ihep.ac.cn}

\abstract{
This work presents the layout design of the \lhcb UP detector, a MAPS-based pixel tracker composed of four detection planes, and several approaches for its standalone track reconstruction. The dedicated UP tracking algorithms demonstrates that efficient standalone reconstruction can be achieved for \lhcb Upgrade II with high purity, reaching efficiencies close to $98\%$ while maintaining a ghost rate below $4\%$. These results indicate that UP standalone tracks can provide high-quality inputs for global \lhcb reconstruction and offer a viable solution for future high-luminosity tracking and real-time reconstruction challenges.}


\begin{document}

\maketitle
\flushbottom

\section{Introduction}\label{sec:intro}

The \lhcb detector \cite{LHCb:2008vvz} is a forward single-arm spectrometer at the \lhc, designed for precision studies of heavy-flavor physics within a unique pseudo-rapidity region of $2<\eta<5$. Since 2022, \lhcb has been operating with the Upgrade~I detector at an instantaneous luminosity of $2\times10^{33}\cm^{-2}\sec^{-1}$, corresponding to five times the original design value \cite{LHCb:2012doh}. The upgraded tracking system consists of the VErtex LOcator (VELO) \cite{Bediaga:2013tje}, the Upstream Tracker (UT), and the Scintillating Fiber Tracker (SciFi) \cite{LHCb:2014uqj}.

To fully exploit the physics potential of the high-luminosity \lhc (HL-LHC), a second major upgrade of the \lhcb detector (Upgrade~II) is planned for the 2030s \cite{LHCb:2021glh}. The experiment is expected to operate at an instantaneous luminosity of up to $1.0\times10^{34}\,\mathrm{cm^{-2}s^{-1}}$, leading to substantially higher detector occupancy and data rates. These conditions will pose significant challenges for track reconstruction, particularly for pattern recognition in dense environments.

All sub-systems of the tracking system, including UT, are planned to be refurbished or replaced during Upgrade II \cite{LHCb:2021glh}. The current silicon-strip-based UT will be replaced by a four-plane pixel detector, the Upstream Pixel Tracker (UP). The UP detector provides measurements upstream of the magnet, improving the association between upstream and downstream track segments as well as the momentum resolution, particularly in the low-\pt region. It is also important for reconstructing displaced tracks from long-lived particles such as \Kz and \Lz.

\section{Design of LHCb UP detector}\label{sec:design}

\begin{figure}[htbp]
    \centering
    \includegraphics[height=0.4\textwidth]{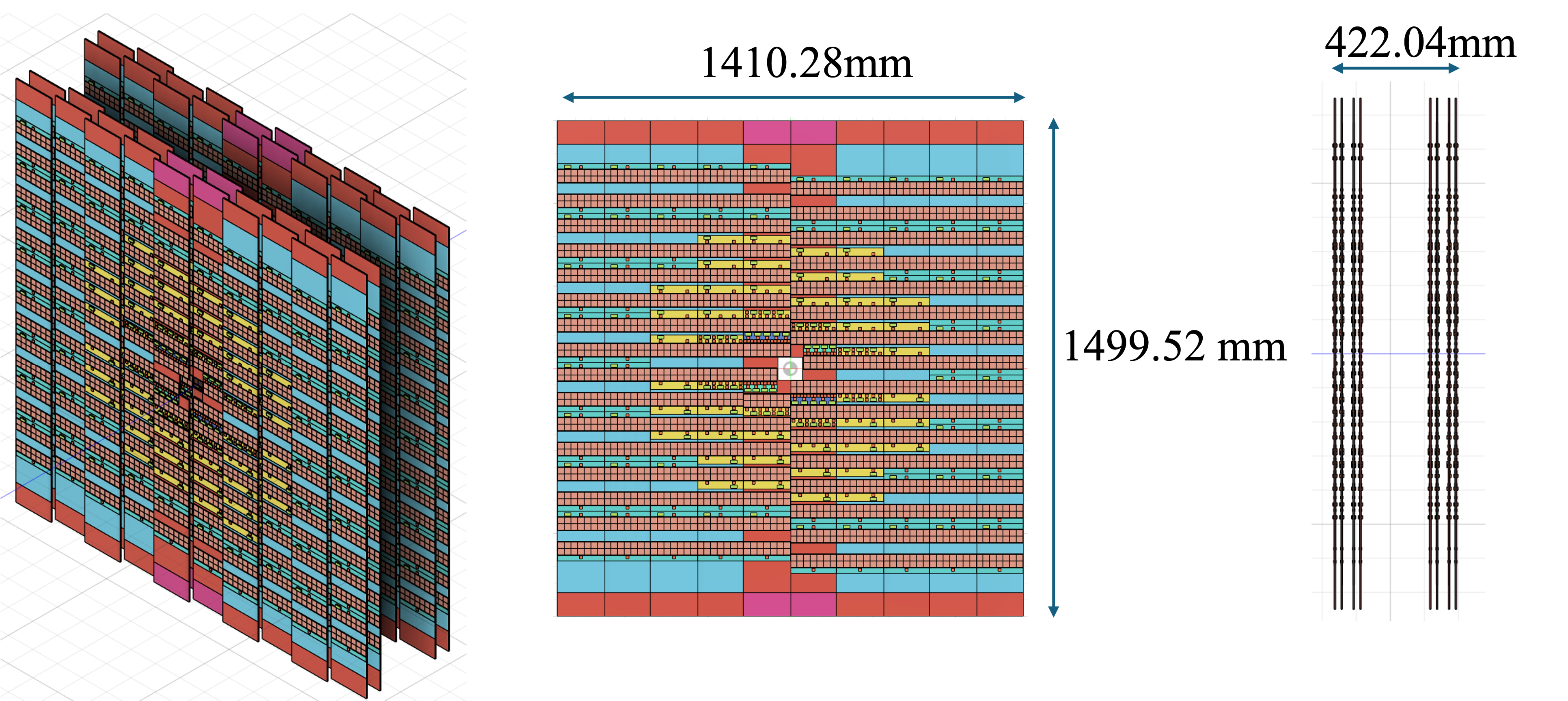}\hspace{1cm}
    \caption{(Left) The overall layout design of the UP detector. (middle) Front cross-sectional view looking perpendicular to the detection plane. (right) Side view along the horizontal direction, highlighting the 4-layer structure with staggered staves implemented for full geometric coverage.
    }
    \label{fig:UP_layout_design}
\end{figure}

\begin{figure}[htbp]
    \centering
    \includegraphics[height=0.4\textwidth]{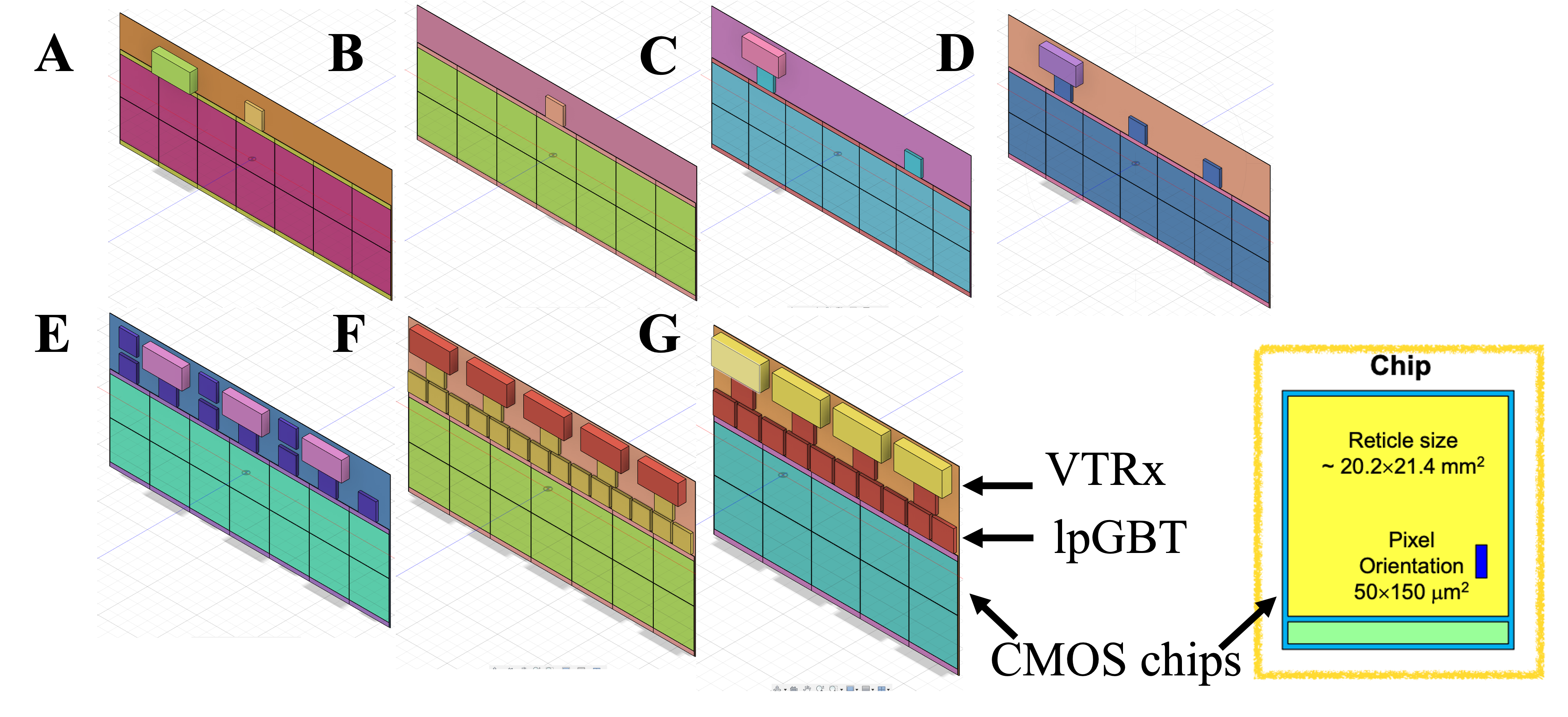}
    \caption{Layout variants of the module designs for the \lhcb UP detector following the specifications documented in the UP FTDR \cite{LHCb:2012doh}. Seven distinct module configurations (labelled A–G) are illustrated, alongside a detailed drawing of the adopted CMOS sensor chip.}
    \label{fig:UP_Module_design}
\end{figure}

The baseline design of the UP detector is illustrated in Figure~\ref{fig:UP_layout_design} and \ref{fig:UP_Module_design}, which follows a structure similar to the UT, consisting of four detector planes. Each UP plane covers an active area of approximately $1400\mm$ in the horizontal direction and $1200\mm$ in the vertical direction. A central opening is reserved for the beam pipe, corresponding to roughly $78\times74\mm^2$ \cite{LHCb:2021glh}. Each UP plane comprises of ten staves and a total 32 modules per stave are mounted alternately on both sides of lightweight support staves. 

The UP module is assembled by interconnecting 14 MAPS pixel detection chips arranged in a $7\times2$ configuration on a flexible circuit. Each chip has dimensions of approximately $2\times2\cm^2$, with a pixel size of $50 \times 150\mum^2$. A guard ring of about $0.8\mm$ surrounds each chip, resulting in small inactive regions between neighboring chips. The average material budget in the UP volume in the range $2<\eta<4.5$ is estimated as $1.2\%~X_0$ for one UP plane \cite{Li:2022spf,Liu:2024xdf}.

\section{UP standalone track reconstruction algorithm}\label{sec:UPstandalone}

Efficient standalone track reconstruction in the UP detector can improve the allocation of computing resources in the \lhcb tracking sequence, optimize the reconstruction speed, and provide additional handles for detector alignment and validation. It also opens new possibilities for maintaining tracking performance under the high data-rate conditions expected in \lhcb Upgrade II. The UT detector is based on silicon strip sensors with strip lengths of approximately $10\cm$ and consists of only four detection planes, making standalone track reconstruction impractical. In contrast, by using bi-dimensional pixel sensors, the UP detector enables standalone track reconstruction, even if retaining the same four-layer geometry.

\begin{figure}[htbp]
    \centering
    \includegraphics[width = 0.35\textwidth]{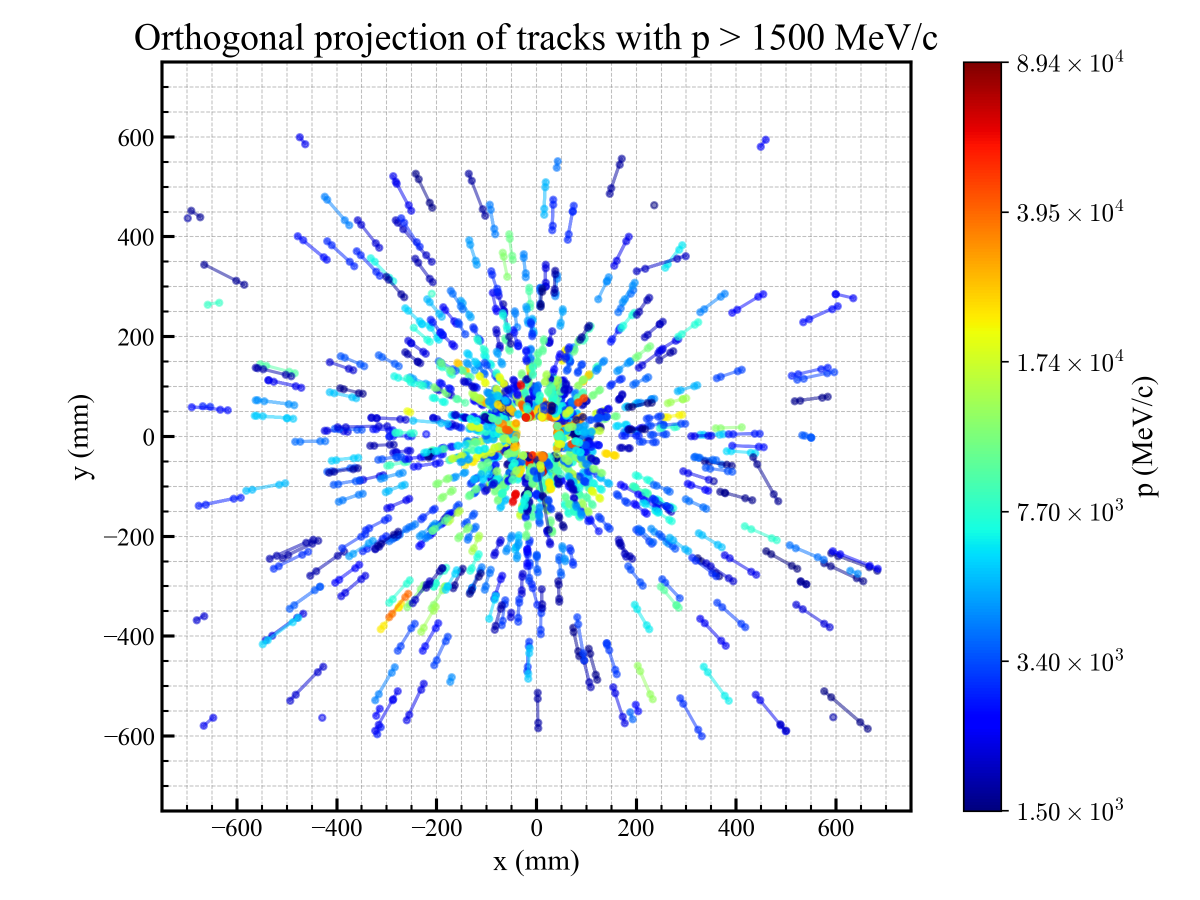}
    \includegraphics[width = 0.35\textwidth]{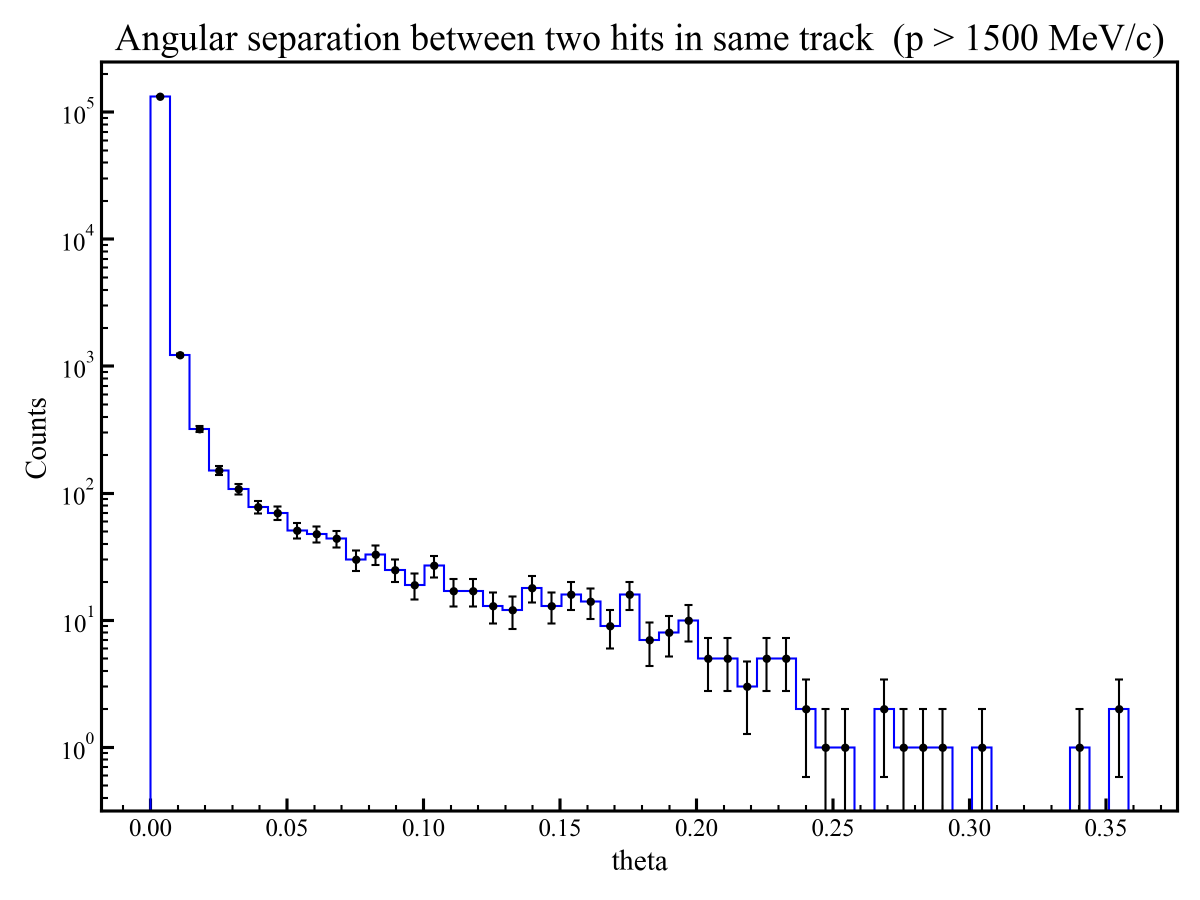}
    \caption{(Left) UP hits (in dots) of non-electron particles and their UP track segments (in lines), projected on the central plane of the UP system. Colors indicate different momentum, while the central empty region corresponds to the beam hole. (Right) Distribution of the angular deviation between the incoming and outgoing directions of non-electron charged particles ($\p>1.5\gevc$) traversing the UP detector within its acceptance.}
    \label{fig:up_track}
\end{figure}

Figure \ref{fig:up_track} shows that most UP tracks are approximately straight lines. Therefore, the UP standalone tracking algorithm can follow the seed-forward strategy used in the VELO tracking \cite{LHCb:2023hlw}. First, seed hits are selected from a reference layer. For each seed, a next-layer search window is opened to identify compatible hit pairs. These hit pairs are then extrapolated to the remaining layers, where the nearest hits within the corresponding search windows are evaluated. If a hit satisfies the selection criteria, it is added to the track segment. An iterative algorithm is introduced, and the search window size is optimized to improve reconstruction efficiency while suppressing ghost tracks. In addition, overlaps between UP modules can introduce extra hits, which are usually used in the reconstruction of ghost tracks. To mitigate this effect, a hit-sharing optimization is applied: when a hit is compatible with multiple track candidates, it is preferentially assigned to the candidate with more matched hits, which is more likely to correspond to a real track.

\begin{figure}[htbp]
    \centering
    \includegraphics[width=0.45\textwidth]{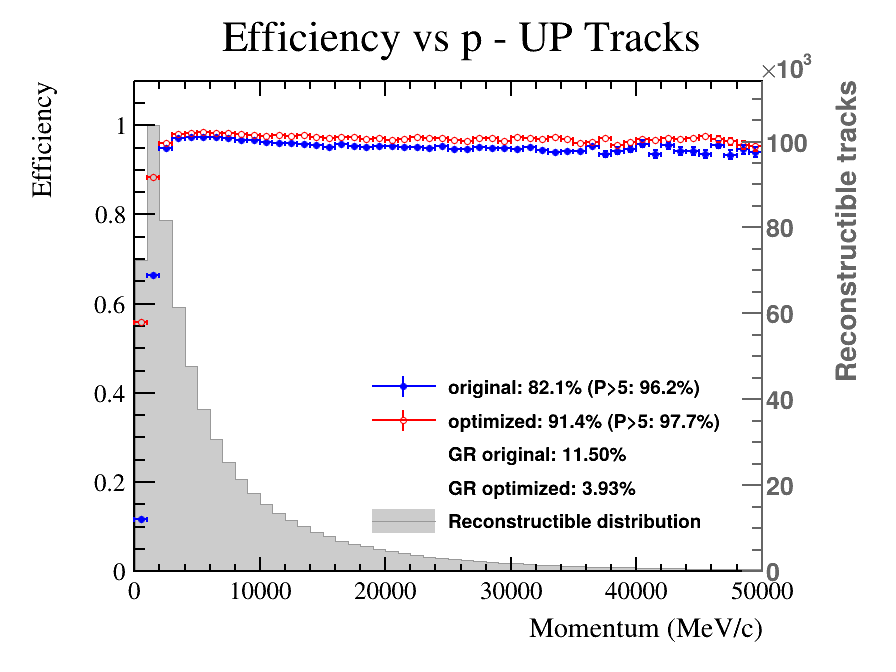}
    \includegraphics[width=0.45\textwidth]{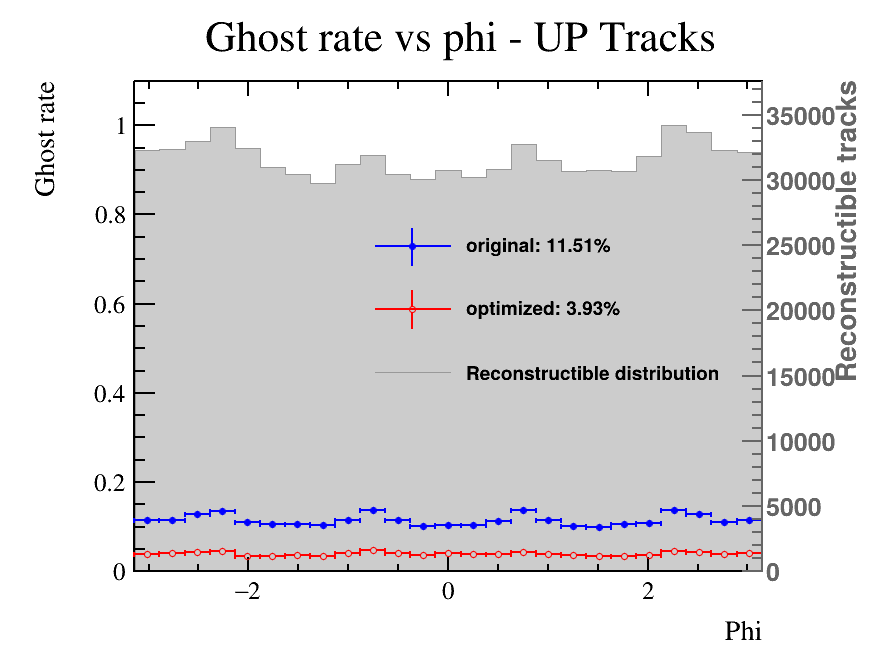}
    \caption{Performance of the UP standalone track reconstruction. The reconstruction efficiency (left) and the ghost track rate (right) based on the original seed-forward algorithm are shown in blue lines and the hit-sharing optimization in red lines. The distributions for the MC particles are the gray area.}
    \label{fig:UPStandalone_performance}
\end{figure}

The performance of the UP standalone track reconstruction is shown in Figure~\ref{fig:UPStandalone_performance}. The efficiency and ghost rate are defined as
\begin{eqnarray}
{\rm Efficiency} &=& {\rm Number~of~reconstructed~tracks~matched~with~MC~truth}\over{\rm Number~of~reconstructible~tracks}\\
{\rm Ghost~rate} &=& {\rm Number~of~reconstructed~tracks~not~matched~or~tracks~no~match~to~MC~truth}\over{\rm Number~of~reconstructed~tracks}
\end{eqnarray}
The overall track reconstruction efficiency by the pattern recognition is $82.1\%$ for all UP standalone tracks, and $96.2\%$ for tracks with $\p>5\gevc$, while the ghost rate is $11.5\%$. After the hit-sharing optimization, the performance is significantly improved, where the efficiency enhanced to $91.4\%$ for all tracks and $97.7\%$ for high momentum ones, and the ghost rate is reduced to $3.9\%$. The resulting high-purity UP tracks provide robust inputs for the global \lhcb tracking reconstruction.

\section{FPGA-based UP standalone track reconstruction algorithm}

The Upgrade II online data throughput is expected to reach about 200\tbps, posing significant challenges for real-time event reconstruction and triggering. To cope with the extremely high data rate, it is essential to explore highly parallelized tracking algorithms optimized for dedicated computing architectures such as FPGAs. In this context, FPGA-based implementations of the UP tracking algorithms are investigated.

During Upgrade I, \lhcb developed the RETINA algorithm for real-time track reconstruction for the VELO detector \cite{Ristori:2000vg,Abba:2014iga,LHCb:2022anl}. Building on this approach, the algorithm is extended to the four-plane UP detector. Benefiting from the intrinsically parallel and pipelined architecture of FPGA devices, the implementation will enable low-latency and high-throughput track reconstruction, providing a promising solution for real-time tracking in the high-occupancy environment.

\begin{figure}[htbp]
    \centering
    \includegraphics[width = 0.37\textwidth]{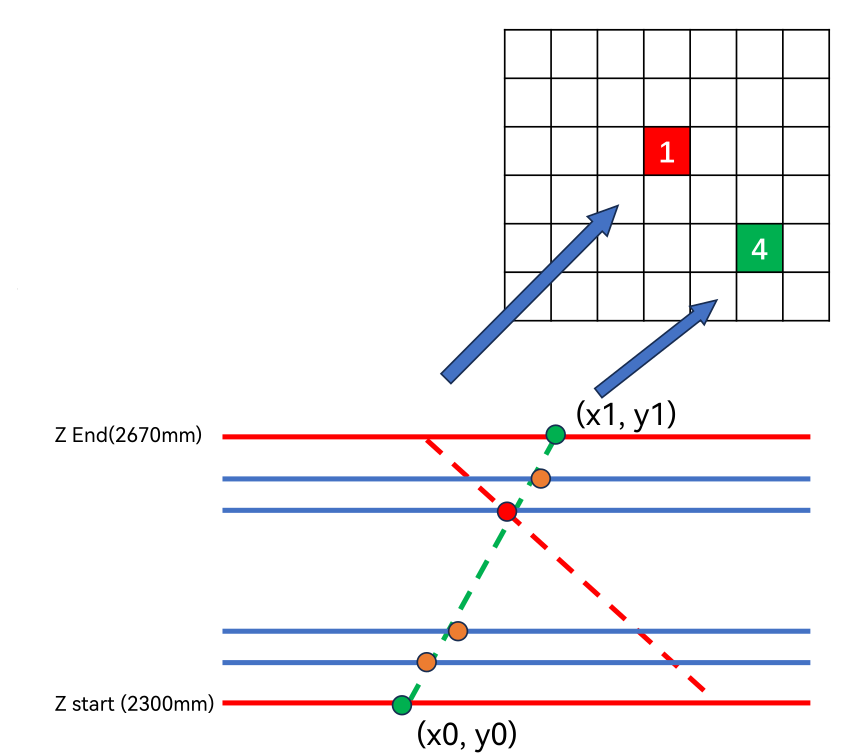}
    \includegraphics[width = 0.32\textwidth]{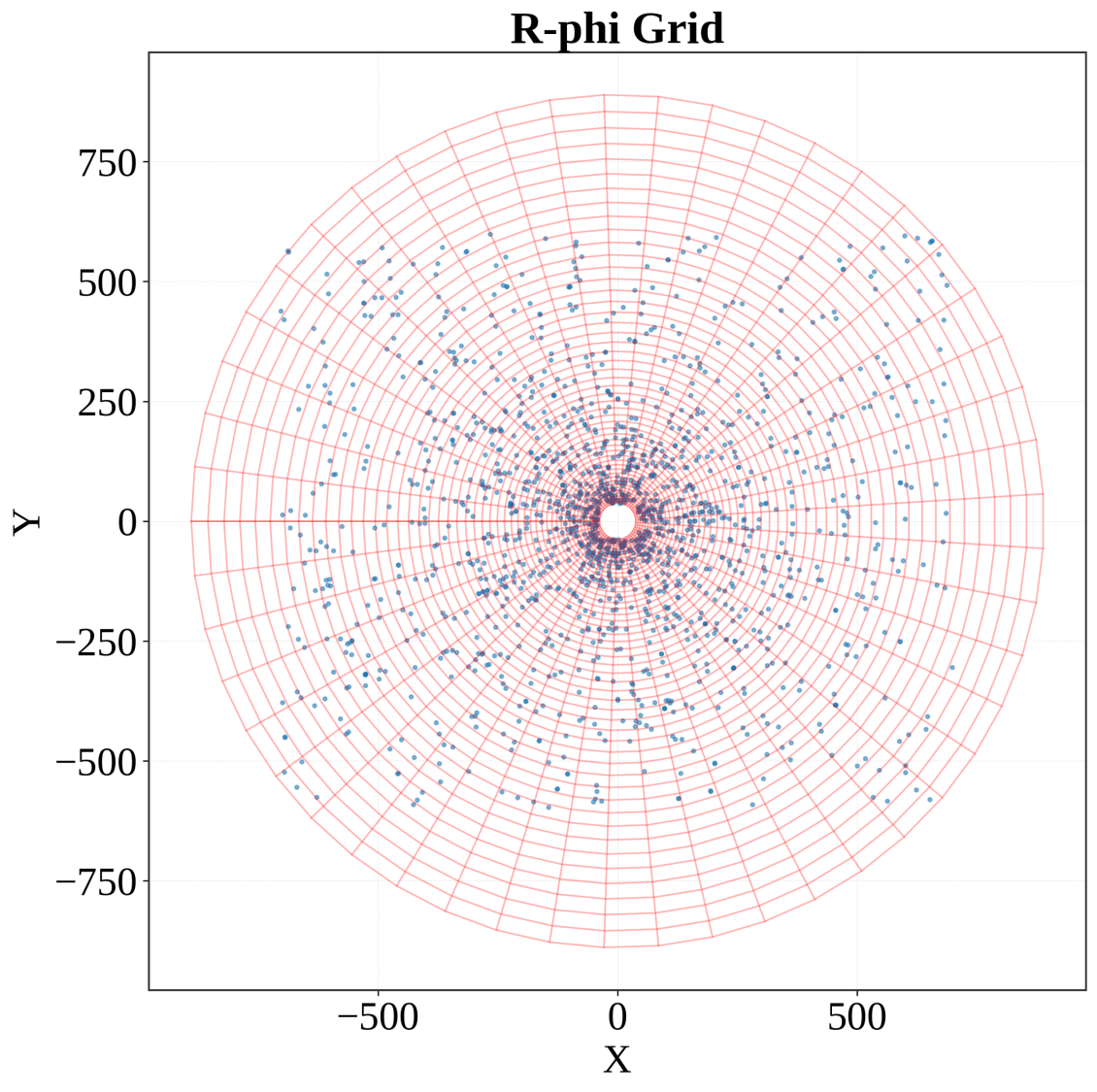}
    \caption{(Left) Schematic representation of the RETINA algorithm. $(x_0,y_0)$ and $(x_1,y_1)$ denote arbitrary grid points in the parameter space defined on two reference planes. The points correspond to hits in the UP detector associated with the same charged-particle track. (Right) Distribution of UP track hit positions on the plane perpendicular to the beam direction. The red grid points represent the bins in the $R-\phi$ parameter space obtained through the Hough transformation.}
    \label{fig:FPGA_algo}
\end{figure}

As illustrated in Figure~\ref{fig:FPGA_algo}, the algorithm is based on the Hough transformation, which maps the track-finding problem to a peak-finding problem in the parameter space. Two reference planes are defined at the entrance and exit of the UP detector. Each plane is parameterized by $(r,\phi)$, leading to a four-dimensional Hough space $(r_0, \phi_0, r_1, \phi_1)$. The parameter space is discretized into $N$ bins per dimension (with $N=150$ in the current design), corresponding to a maximum of $150^4\sim500{\rm M}$ cells. However, in practice, only a small subset of cells is populated after applying loose phase-space constraints, reducing the number of active cells to approximately 6M. Detector hits are projected into this Hough space, where they increment the corresponding cells. Track candidates are identified as cells containing at least three hits. Peaks in the Hough space are then selected and refined using a \chisq-based linear regression, yielding the final UP track segments.

\begin{figure}[htbp]
    \centering
    \includegraphics[height=0.3\textwidth]{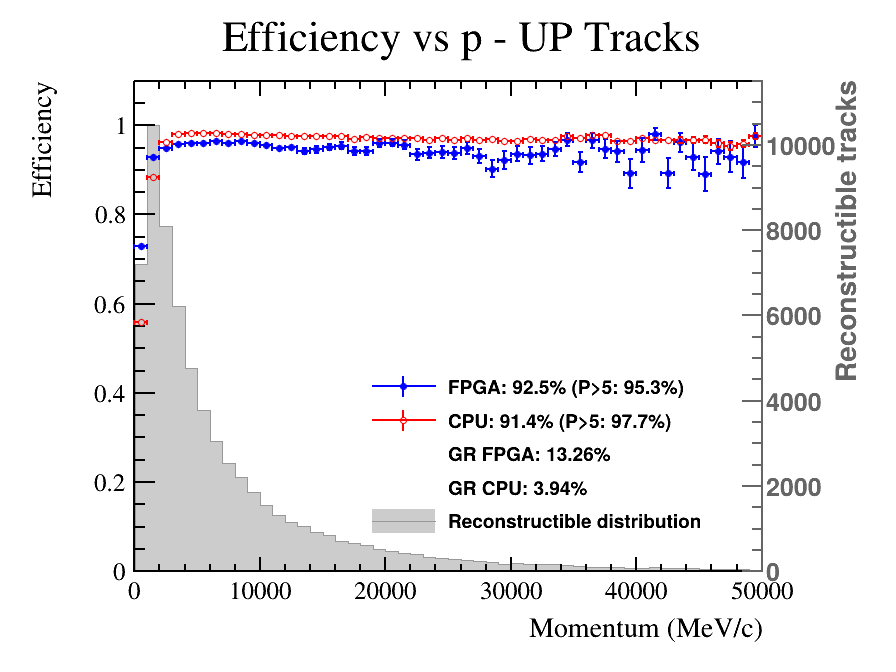}
    \caption{Comparison of the UP standalone track reconstruction efficiencies as functions of the momentum, derived via the RETINA algorithm (blue line) and the seed-forward algorithm (red line).}
    \label{fig:UPStandalone_comparison}
\end{figure}

Figure~\ref{fig:UPStandalone_comparison} shows the reconstruction performance of the FPGA-based RETINA algorithm developed for UP standalone tracking, compared with the aforementioned seed-forward algorithm. The RETINA algorithm achieves a tracking performance comparable to that of the conventional seed-forward approach, particularly in terms of the overall reconstruction efficiency. In the low-momentum region, a noticeable improvement in reconstruction efficiency is observed with the RETINA-based method. This behavior is attributed to the Hough-transform-based reconstruction strategy, which performs pattern recognition in the track-parameter space and is therefore well suited to the trajectory topology in the \lhcb detector. Furthermore, the RETINA implementation includes a track-fit stage for candidate refinement, while the current seed-forward algorithm is based solely on pattern recognition. Conversely, the FPGA implementation exhibits a slightly higher ghost rate and a somewhat lower efficiency in the high-momentum region. The seed-forward algorithm has undergone extensive optimization, including iterative reconstruction and hit-sharing treatments to suppress redundant hits from detector overlaps, resulting in improved efficiency and reduced ghost rates. These optimizations have not yet been fully incorporated into the present RETINA implementation. Therefore, the observed differences are mainly attributed to the current stage of algorithm development rather than to intrinsic limitations of the RETINA approach.

\section{Summary}

This work presents standalone tracking and FPGA-based reconstruction for the UP detector in the \lhcb Upgrade II. A seed-forward algorithm with hit-sharing optimization achieves about $91.4\%$ overall efficiency ($~97.7\%$ for high-momentum tracks) with a ghost rate below $3.9\%$, demonstrating high-purity track segment reconstruction in a dense environment. In parallel, an FPGA-oriented implementation based on a RETINA approach achieves comparable performance to the software algorithm while enabling low-latency, highly parallel processing.

These results demonstrate that UP-standalone tracking is a viable and robust component for the Upgrade II real-time tracking system. The approach provides a promising foundation for future \lhcb real-time reconstruction at even higher luminosity, where fast pattern recognition and hardware-accelerated tracking will become increasingly critical.

\acknowledgments
This work was partially supported by the National Key Research and Development Program of China under Grant number 2023YFA1606300, the National Natural Science Foundation of China (NSFC) under Grant numbers W2443008 and 12188102, the science and technology innovation project of Institute of High Energy Physics Chinese Academy of Sciences with the fund number 2024000077, and the High Energy Physics Research Center of Henan Academy of Sciences.





\end{document}